\documentclass[manuscript]{aastex}

\shorttitle{Astrometric Microlensing by the Ellis Wormhole}
\shortauthors{Y. Toki, T. Kitamura, H. Asada, F. Abe}

\begin{document}


\title{Astrometric Image Centroid Displacements due to 
Gravitational Microlensing by the Ellis Wormhole}


\author{Yukiharu Toki, Takao Kitamura, Hideki Asada}
\affil{Faculty of Science and Technology, 
Hirosaki University, Hirosaki 036-8561, Japan\\
}
\and 
\author{Fumio Abe\altaffilmark{1}}
\affil{Solar-Terrestrial Environment Laboratory, Nagoya University \\
Furo-cho, Chikusa-ku, Nagoya 464-8601, Japan \\
abe@stelab.nagoya-u.ac.jp}


\altaffiltext{1}{Nagoya University Southern Observatories}


\begin{abstract}
Continuing work initiated in an earlier publication 
(Abe, ApJ, {\bf 725} (2010) 787), 
we study the gravitational microlensing 
effects of the Ellis wormhole in the weak-field limit. 
First, we find a suitable coordinate transformation, 
such that the lens equation 
and analytic expressions of the lensed image positions 
can become much simpler than the previous ones. 
Second, we prove that two images always appear 
for the weak-field lens by the Ellis wormhole. 
By using these analytic results, 
we discuss astrometric image centroid displacements due to 
gravitational microlensing by the Ellis wormhole. 
The astrometric image centroid trajectory 
by the Ellis wormhole is different from the standard one 
by a spherical lensing object that is expressed by 
the Schwarzschild metric. 
The anomalous shift of the image centroid by the Ellis wormhole lens 
is smaller than that by the Schwarzschild lens, 
provided that the impact parameter and the Einstein ring radius 
are the same. 
Therefore, the lensed image centroid by the Ellis wormhole 
moves slower. 
Such a difference, though it is very small, will be 
in principle applicable for detecting or constraining 
the Ellis wormhole by using future high-precision astrometry observations.  
In particular, the image centroid position gives us an additional
information, so that the parameter degeneracy existing 
in photometric microlensing can be partially broken. 
The anomalous shift reaches the order of a few micro arcsec. 
if our galaxy hosts a wormhole with throat radius larger than $10^5$ km. 
When the source moves tangentially to the Einstein ring for instance, 
the maximum position shift of the image centroid 
by the Ellis wormhole is  
$0.18$ normalized by the Einstein ring radius. 
For the same source trajectory, 
the maximum difference between 
the centroid displacement by the Ellis wormhole lens 
and that by the Schwarzschild one 
with the same Einstein ring radius is $-0.16$ 
in the units of the Einstein radius, 
where the negative means that the astrometric displacement 
by the Ellis wormhole lens is smaller than 
that by the Schwarzschild one. 
\end{abstract}




\section{Introduction}
A peculiar feature of general relativity is that the theory 
admits a nontrivial topology of a spacetime. 
A solution of the Einstein equation that connects distant points 
of space--time was introduced by \citet{ein35}. 
This solution called the {\it Einstein--Rosen bridge} 
was the first solution to later be referred to as a wormhole. 
Initially, this type of solution was just a trivial or 
teaching example of mathematical physics. 
However, \citet{morr88} proved that some wormholes are "traversable"; 
i.e., space and time travel can be achieved by passing 
through the wormholes. 
They also showed that the existence of a wormhole requires 
exotic matter that violates the null energy condition. 
The existence of wormholes, though they are very exotic, 
has not been ruled out in theory. 
Inspired by the Morris--Thorne paper, a number of theoretical works 
(see \citet{vis95, lob09} and references therein) on wormholes 
have been done. 
The curious nature of wormholes such as time travel, 
negative energy conditions, space--time foams, and 
growth of a wormhole in an accelerating universe have been studied. 
In spite of enthusiastic theoretical investigations, 
studies searching for observational evidence of the existence 
of wormholes are scarce. 
Only a few attempts have been made to show the existence 
or nonexistence of wormholes in our universe 
\citep{cram95, tor98, saf01, bog08}. 

A possible observational method that has been proposed to detect 
or exclude the existence of wormholes is the application 
of optical gravitational lensing, since the light ray propagation 
is sensitive to a local spacetime geometry.  
The gravitational lensing of wormholes was pioneered 
by \citet{cram95}, who inferred that some wormholes show 
"negative mass" lensing. 
They showed that the light curve of the negative-mass lensing event 
of a distant star has singular double peaks. 
Several authors subsequently conducted theoretical studies 
on detectability \citep{saf01, bog08}. 
Another gravitational lensing method employing gamma rays was 
proposed by \citet{tor98}, who postulated that the singular 
negative-mass lensing of distant active galactic nuclei causes 
a sharp spike of gamma rays and may be observed 
as double-peaked gamma-ray bursts.  
They analyzed BATSE data to put an upper limit to 
the negative mass density $O(10^{-36}) \:\mbox{g cm}^{-3}$
in the form of wormholelike objects. 

There have been several recent works 
\citep{sh04, per04, nan06, rah07, dey08, abe10, asada11} 
on the gravitational lensing of wormholes as structures of
space--time. 
Such studies are expected to unveil lensing properties 
directly from the space--time structure. 
One study \citet{dey08} calculated the deflection angle of light due
to the Ellis wormhole, whose asymptotic mass at infinity is zero. 
The massless wormhole is particularly interesting 
because it is expected to have unique gravitational lensing effects.
This type of wormhole was first introduced by \cite{ell73} 
as a massless scalar field. 
Later, \citet{morr88} studied this wormhole and proved it 
to be traversable. 
The dynamical feature was studied by \cite{shi02}, 
who showed that Gaussian perturbation causes either explode 
to an inflationary universe or collapse to a black hole. 
\cite{das05} showed that the tachyon condensate can be a source 
for the Ellis geometry. 
\citet{abe10} provided a method to calculate light curves 
of the gravitational microlensing of the Ellis wormhole 
in the weak-field limit. 
This result has been discussed as one example of 
corrections to the standard formula of the deflection angle 
by \cite{asada11}. 

The main results of this paper are:  
(1) We derive analytic expressions for 
calculations of the wormhole lensing. 
(2) We show the astrometric image centroid trajectory 
by the Ellis wormhole lens. 
Studies of centroid displacements of lensed images 
have been limited within the Schwarzschild lens 
\citep{Walker,MY,HOF,SDR,JHP,asada02,HL}. 
In Section 2, we discuss gravitational lensing by the Ellis wormhole
in the weak-field limit. 
We use a suitable coordinate transformation, 
such that the lens equation to determine image positions 
can take a much simpler form than the previous one derived by 
\citet{abe10}. 
By using this method, calculations of the gravitational microlensing 
by the Ellis wormhole are shortened. 
In addition, we prove that two images always appear 
for the weak-field lens by the Ellis wormhole. 
In section 3, we discuss astrometric image centroid displacements due to 
gravitational microlensing by the Ellis wormhole. 
The results are summarized in Section 4.

\section{Gravitational lensing by the Ellis wormhole: 
Weak-field approximation} 
\subsection{Ellis wormhole lens}
Magnification of the apparent brightness of a distant star by the
gravitational lensing effect of another star was predicted by
\citet{ein36}. This kind of lensing effect is called "microlensing"
because the images produced by the gravitational lensing are very
close to each other and are difficult for the observer to resolve. 
The observable effects are not only 
the changing apparent brightness of the source star 
but also the shift in the image centroid position. 
The brightness changing effect 
was discovered in 1993 
\citep{uda93, alc93, aub93} and has been used to detect 
astronomical objects that do not emit observable signals 
(such as visible light, radio waves, and X rays) or 
are too faint to observe. 
Microlensing has successfully been applied to detect 
extrasolar planets \citep{bon04, bea06} and brown dwarfs 
\citep{nov09, gou09}. 
Microlensing is also used to search for unseen black holes 
\citep{alc01, ben02, poi05} and massive compact halo objects 
\citep{alc00, tis07, wyr09}, a candidate for dark matter.

\begin{figure}
\epsscale{.80}
\plotone{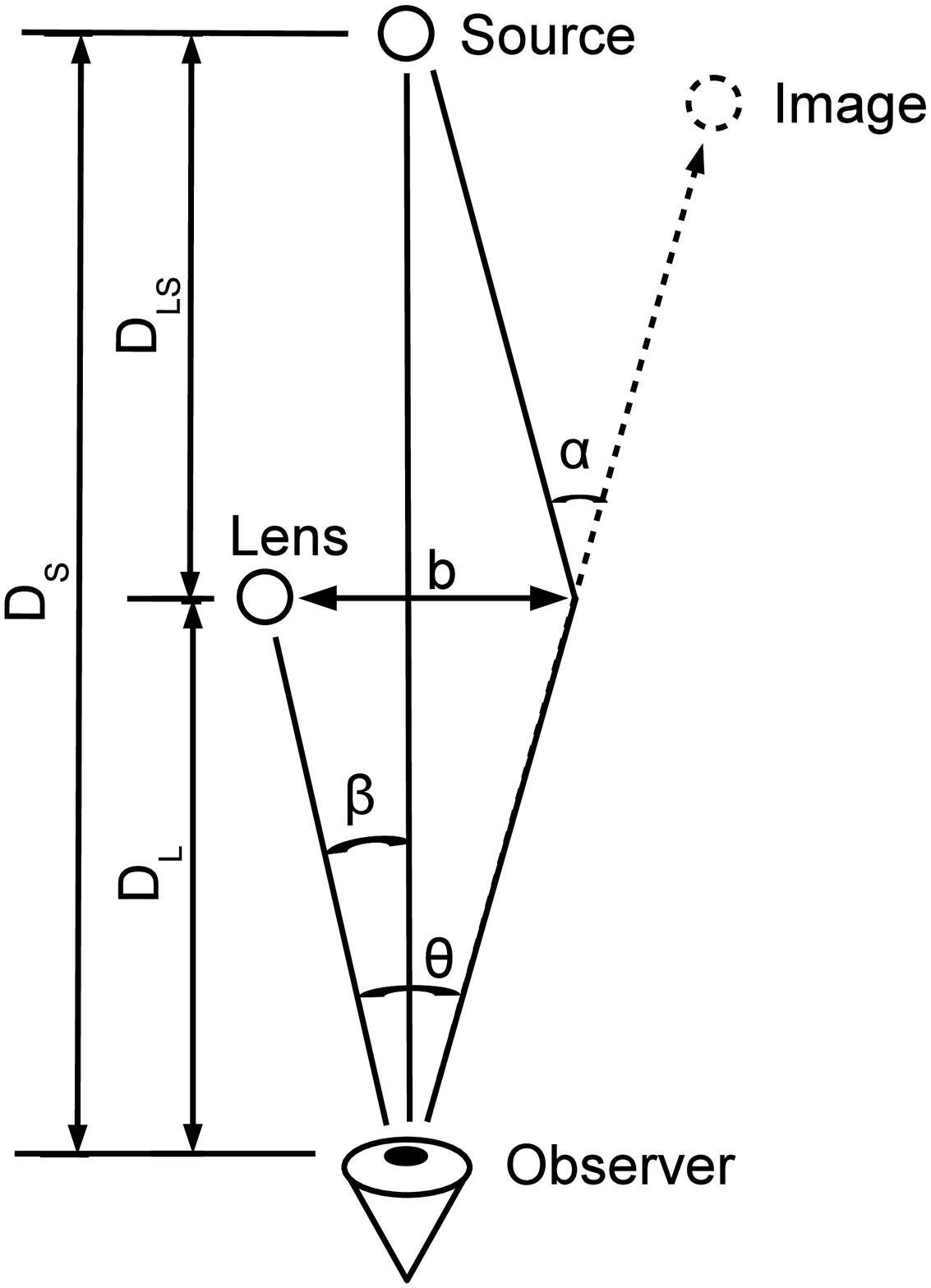}
\caption{Sketch of the relation between the source star, 
lens (wormhole), and observer. \label{fig1}}
\end{figure}

The gravity of a star is well expressed by the Schwarzschild
metric. The gravitational microlensing of the Schwarzschild metric
\citep{ref64, lieb64, Pacz86} has been studied in the weak-field
limit. In this section, we simply follow the method used for
Schwarzschild lensing. Figure \ref{fig1} shows the relation between
the source star, the lens (wormhole), and the observer. The Ellis
wormhole is known to be a massless wormhole, which means that the
asymptotic mass at infinity is zero. 
The Ellis wormhole is expressed by the line element
\begin{equation}
ds^2 = dt^2 - dr^2 - (r^2 + a^2) (d\theta^2 + sin^2(\theta) d\phi^2),
\end{equation}
where $a$ is the throat radius of the wormhole. 
However, this wormhole deflects light by gravitational 
lensing \citep{cle84,che84,nan06,dey08} because of 
its curved space--time structure. 

The deflection angle $\alpha(r)$ of the Ellis wormhole 
was derived by \cite{dey08} to be
\begin{equation}
\alpha(r) = \pi \left\{\sqrt \frac{2 (r^2 + a^2)}{2 r^2 + a^2} -1 \right\},
\end{equation}
where $r$ is the closest approach of the light.
In the weak-field limit ($r \rightarrow \infty$), the deflection angle becomes
\begin{equation}
\alpha(r) \rightarrow \frac{\pi}{4}\frac{a^2}{r^2} 
+ o\left(\frac{a}{r}\right)^4. \label{eqn:defl}
\end{equation}
Note that 
\cite{dey08} treatment is true of the weak-field region 
but it may be corrected in the strong-field one, 
because they assume that $r=0$ were singular and it 
could be excluded as is the case for the Schwarzschild metric. 
For the Ellis wormhole, however, $r=0$ is not a singularity 
but a regular sphere and hence 
$r=0$ cannot be excluded in the strong-field lensing. 

The angle between the lens (wormhole) and the source $\beta$ can then be written as
\begin{equation}
\beta  = \frac{1}{D_L} b - \frac{D_{LS}}{D_S}\alpha(r),
\end{equation}
where $D_L$, $D_S$, $D_{LS}$, and $b$ are the distances from the observer to the lens, from the observer to the source, and from the lens to the source, and the impact parameter of the light, respectively. In the asymptotic limit, Schwarzschild lensing and massive Janis--Newman--Winnicour (JNW) wormhole lensing \citep{dey08} have the same leading term of $o\left(1/r\right)$. Therefore, the lensing property of the JNW wormhole is approximately the same as that of Schwarzschild lensing and is difficult to distinguish. As shown in Equation (\ref{eqn:defl}), the deflection angle of the Ellis wormhole does not have the term of $o\left(1/r\right)$ and starts from $o\left(1/r^2\right)$. This is due to the massless nature of the Ellis wormhole and indicates the possibility of observational discrimination from the ordinary gravitational lensing effect. In the weak-field limit, $b$ is approximately equal to the closest approach $r$. For the Ellis wormhole, $b = \sqrt{r^2 + a^2} \rightarrow r (r \rightarrow \infty)$. We thus obtain
\begin{equation}
\beta = \frac{r}{D_L} - \frac{\pi}{4} \frac{D_{LS}}{D_S}\frac{a^2}{r^2}\hspace{1cm}(r > 0). \label{eqn:proj} 
\end{equation}
 The light passing through the other side of the lens may also form images. However, Equation (\ref{eqn:proj}) represents deflection in the wrong direction at $r < 0$. Thus, we must change the sign of the deflection angle: 
\begin{equation}
\beta = \frac{r}{D_L} + \frac{\pi}{4} \frac{D_{LS}}{D_S}\frac{a^2}{r^2}\hspace{1cm}(r < 0). 
\label{eqn:proj2} 
\end{equation}
It would be useful to note that a single equation is suitable both for $r > 0$ and $r < 0$ images in the Schwarzschild lensing. However, such treatment is applicable only when the deflection angle is an odd function of $r$. 
 
 If the source and lens are completely aligned along the line of sight, the image is expected to be circular (an Einstein ring). The Einstein radius $R_E$, which is defined as the radius of the circular image on the lens plane, is obtained from Equation (\ref{eqn:proj}) with $\beta = 0$ as
 \begin{equation}
 R_E = \sqrt[3]{\frac{\pi}{4}\frac{D_L D_{LS}}{D_S} a^2}. \label{eqn:re}
 \end{equation}
 
The image positions can then be calculated from
\begin{equation}
\beta = \theta - \frac{\theta_E^3}{\theta^2} \hspace{1cm} (\theta > 0)  \label{eqn:imgeq}
\end{equation}
and 
\begin{equation}
\beta = \theta + \frac{\theta_E^3}{\theta^2} \hspace{1cm} (\theta < 0),  \label{eqn:imgeq2}
\end{equation}
where $\theta = b / D_L  \thickapprox r / D_L$ is the angle between the image and lens, and $\theta_E = R_E / D_L$ is the angular Einstein radius. Using reduced parameters $\hat{\beta} = \beta / \theta_E$ and $\hat{\theta} = \theta / \theta_E$, Equations (\ref{eqn:imgeq}) and (\ref{eqn:imgeq2}) become simple cubic formulas: 
\begin{equation}
\hat{\theta}^3 - \hat{\beta} \hat{\theta}^2 -1 = 0 \hspace{1cm} (\hat{\theta} > 0) \label{eqn:poly}
\end{equation}
and
\begin{equation}
\hat{\theta}^3 - \hat{\beta} \hat{\theta}^2 +1 = 0 \hspace{1cm} (\hat{\theta} < 0). \label{eqn:poly2}
\end{equation}

Following Abe (2010), first, we briefly summarize how to obtain 
the roots of the above equations. 
 As the discriminant of Equation (\ref{eqn:poly}) is $-4\hat{\beta}^3 - 27 < 0$, Equation (\ref{eqn:poly}) has two conjugate complex solutions and a real solution:
 \begin{equation}
 \hat{\theta} = \frac{\hat{\beta}}{3} + U_{1+} + U_{1-} ,
\label{theta+}
 \end{equation} 
 with, 
 \begin{equation}
 U_{1\pm} = \sqrt[3]{\frac{\hat{\beta}^3}{27} + \frac{1}{2} \pm \sqrt{\frac{1}{4}\left(1 + \frac{2 \hat{\beta}^3}{27}\right)^2 - \frac{\hat{\beta}^6}{27^2}}}.
 \end{equation}
The real positive solution corresponds to the physical image. 

The discriminant of Equation (\ref{eqn:poly2}) is $4\hat{\beta}^3 - 27$. Thus it has a real solution if $\hat{\beta} < \sqrt[3]{27/4}$: 
\begin{equation}
\hat{\theta} = \frac{\hat{\beta}}{3} + U_{2+} + U_{2-}, 
\label{theta-}
\end{equation}
where, 
\begin{equation}
U_{2\pm} = \omega \sqrt[3]{\frac{\hat{\beta}^3}{27} - \frac{1}{2} \pm \sqrt{\frac{1}{4}\left(1 - \frac{2 \hat{\beta}^3}{27}\right)^2 - \frac{\hat{\beta}^6}{27^2}}} , 
\end{equation}
with $\omega \equiv e^{(2\pi/3)i}$. 
This solution corresponds to a physical image inside the Einstein ring. For $\hat{\beta} > \sqrt[3]{27/4}$, Equation (\ref{eqn:poly2}) has three real solutions. However, two of them are not physical because they do not satisfy $\hat{\theta} < 0$. Only the solution 
\begin{equation}
\hat{\theta} = \frac{\hat{\beta}}{3} +  \omega U_{2+} + U_{2-}
\end{equation}
corresponds to a physical image inside the Einstein ring. 

In both cases of $\hat{\theta} > 0$ and $\hat{\theta} < 0$, 
careful treatments of $\hat{\beta}$ inside the square roots 
and the cube ones are required in order to 
know which they are a real positive, real negative or complex.

\subsection{Simplified expressions of lensed image positions} 
Next, let us consider an appropriate coordinate transformation as 
\begin{equation}
u \equiv \frac{1}{\hat{\theta}} , 
\label{u}
\end{equation}
so that Eqs. (\ref{eqn:poly}) and (\ref{eqn:poly2}) can be 
rewritten respectively as 
\begin{equation}
u^3 + \hat{\beta} u -1 = 0 \hspace{1cm} (u > 0) 
\label{eqn:cubic1}
\end{equation}
and
\begin{equation}
u^3 - \hat{\beta} u +1 = 0 \hspace{1cm} (u < 0). 
\label{eqn:cubic2}
\end{equation}
Note that these equations take exactly the standard form 
called {\it depressed cubic} for using Cardano's method 
to find analytic roots of a cubic equation. 
In next subsetion, it is proven that 
there exists the only one true root for each equation. 
Hence, we immediately get the unique real root for $u > 0$ as 
\begin{eqnarray}
u_{1} &=& \frac{1}{\hat{\theta}_{1}} 
\nonumber\\
&=& \sqrt[3]{\frac12 + \sqrt{\frac14 + \frac{\hat{\beta}^3}{27}}} 
- \sqrt[3]{-\frac12 + \sqrt{\frac14 + \frac{\hat{\beta}^3}{27}}} , 
\label{u+}
\end{eqnarray}
and the unique real one for $u < 0$ as 
\begin{eqnarray}
u_{2} &=& \frac{1}{\hat{\theta}_{2}} 
\nonumber\\
&=& - \sqrt[3]{\frac12 + \sqrt{\frac14 - \frac{\hat{\beta}^3}{27}}} 
- \sqrt[3]{\frac12 - \sqrt{\frac14 - \frac{\hat{\beta}^3}{27}}} .  
\label{u-}
\end{eqnarray} 
Clearly it can be shown by direct but lengthy calculations 
that Eqs. (\ref{u+}) and (\ref{u-}) agree with 
Eqs. (\ref{theta+}) and (\ref{theta-}), respectively. 
Note that they are much simpler 
than Eqs. (\ref{theta+}) and (\ref{theta-}). 
In particular, these improved expressions make it 
much easier to see the sign of the argument of 
the square root and the cube one compared with 
Eqs. (\ref{theta+}) and (\ref{theta-}). 
For $\hat{\beta} > \sqrt[3]{27/4}$, 
the argument of the square root in Eq. (\ref{u-}) 
is always negative, so that 
Eq. (\ref{u-}) can be rewritten as 
\begin{eqnarray}
u_{2} 
&=& -2 \sqrt{\frac{\hat{\beta}}{3}} 
\cos \left[ \frac13 
\arctan\left(2 \sqrt{\frac{\hat{\beta}^3}{27} - \frac14}\right) 
\right] .  
\label{u+2}
\end{eqnarray} 
Here we used a relation for two real numbers $p > 0$ and $q < 0$ as 
\begin{eqnarray}
\sqrt[3]{p + \sqrt{q}} + \sqrt[3]{p - \sqrt{q}} 
&=& 
2 \sqrt[3]{r} 
\cos\left[ \frac13 \arctan\left( \frac{\sqrt{-q}}{p} \right)\right] .  
\label{relation}
\end{eqnarray} 
This relation can be shown as follows. 
For $p > 0$ and $q < 0$, we can put 
$p + \sqrt{q}  
= r \exp(i\phi)$ 
by introducing a radial distance defined as 
$r \equiv \sqrt{p^2 - q}$ 
and an angle coordinate defined as 
$\tan\phi \equiv p^{-1} \sqrt{-q}$. 
{}From its complex conjugate we obtain 
$p - \sqrt{q} = r \exp(- i\phi)$. 
Hence we get 
\begin{eqnarray}
\sqrt[3]{p + \sqrt{q}} 
&=& 
\sqrt[3]{r} \exp(i\phi/3) , 
\\
\sqrt[3]{p - \sqrt{q}} 
&=& 
\sqrt[3]{r} \exp(-i\phi/3) , 
\end{eqnarray}
both of which are combined to get Eq. (\ref{relation}).

\subsection{Number of images lensed by the Ellis wormhole} 
It is not convenient to use Cardano's formulas 
to know how many images appear for the Ellis wormhole lens, 
since the formulas include a combination of square roots 
and cube ones 
and therefore straightforward but lengthy calculations 
are required to know the sign of the argument 
of the roots. 

In order to bypass such difficulties, 
we use Descartes' rule of signs (e.g., \cite{Waerden}), 
which states that the number of positive roots either equals 
that of sign changes in coefficients of a polynomial 
(ignoring powers which do not appear) or 
less than it by a multiple of two. 
This theorem tells that Eqs. (\ref{eqn:poly}) and (\ref{eqn:cubic1}) 
have the only one positive root, 
because the sign of the coefficient of each power 
(ignoring powers which do not appear)
is $+$, $-$, $-$ for Eq. (\ref{eqn:poly}) 
and $+$, $+$, $-$ for Eq. (\ref{eqn:cubic1}). 
For Eqs. (\ref{eqn:poly2}) and (\ref{eqn:cubic2}), 
we make a parity transformation as 
$\hat{\theta}^{'} = - \hat{\theta}$ and 
$\hat{u}^{'} = - \hat{u}$, 
so that we can directly apply the Descartes' theorem. 
After the parity transformation is made for 
Eqs. (\ref{eqn:poly2}) and (\ref{eqn:cubic2}), 
the sign of the coefficient of each power is 
$-$, $-$, $+$ for Eq. (\ref{eqn:poly2}) 
and $-$, $+$, $+$ for Eq. (\ref{eqn:cubic2}). 
Therefore, we have the only one negative root 
for each of Eqs. (\ref{eqn:poly2}) and (\ref{eqn:cubic2}). 

The L.H.S. of Eq. (\ref{eqn:cubic1}) becomes $-1$ and $\beta \geq 0$ 
for $u=0$ and $1$, respectively. 
The continuity of the L.H.S. thus means that 
the positive root $u_1$ and $\hat\theta_1$ satisfy 
\begin{eqnarray}
&& 0 < u_1 \leq 1 , 
\\
&& 1 \leq \hat\theta_1 < +\infty ,  
\end{eqnarray}
respectively. 
The L.H.S. of Eq. (\ref{eqn:cubic2}) becomes 
$-\infty$ and $\beta \geq 0$ 
for $u=-\infty$ and $-1$, respectively. 
Hence, the continuity of the L.H.S. means that 
the negative root $u_2$ and $\hat\theta_2$ satisfy 
\begin{eqnarray}
&& u_2 \leq -1 , 
\\
&& -1 \leq \hat\theta_2 < 0 , 
\end{eqnarray}
respectively. 
The above inequalities on $\hat\theta_1$ and $\hat\theta_2$ 
hold also for the Ellis wormhole 
similarly to the Schwarzschild lens.

\section{Astrometric image centroid displacements by the Ellis wormhole}
The light curve of Schwarzschild lensing was derived by
\citet{Pacz86}, whereas  
the counterpart by the Ellis wormhole was calculated by \citet{abe10}.  
The magnification of the brightness for each image 
by the Ellis wormhole lens is 
\begin{eqnarray}
A_1 &\equiv&  
\left|\frac{\hat{\theta}_1}{\hat{\beta}} 
\frac{d\hat{\theta}_1}{d\hat{\beta}}\right| 
\nonumber\\
&=& 
\frac{1}{\left(1 - \frac{1}{\hat{\theta}_1^3}\right) 
\left(1 + \frac{2}{\hat{\theta}_1^3}\right)}
\nonumber\\
&=& 
\frac{1}{(1 - u_1^3) (1 + 2u_1^3)} , 
\label{A1}
\\
A_2&\equiv&
\left|\frac{\hat{\theta}_2}{\hat{\beta}} 
\frac{d\hat{\theta}_2}{d\hat{\beta}}\right|
\nonumber\\
&=& 
\frac{1}{\left(1 + \frac{1}{\hat{\theta}_2^3}\right) 
\left(\frac{2}{\hat{\theta}_2^3} - 1\right)} 
\nonumber\\
&=& 
\frac{1}{(1 + u_2^3) (2u_2^3 - 1)} , 
\label{A2}
\end{eqnarray} 
where $A_1$ and $A_2$ are magnification of the outer and inner images,  
$\hat{\theta}_1$ and $\hat{\theta}_2$ correspond to 
outer and inner images, respectively. 
Here, we use 
$0< u_1 \leq 1$and $u_2 \leq -1$. 
Hence, the total magnification of the brightness $A$ is
 \begin{eqnarray}
 A  &\equiv& A_1 + A_2 
\nonumber\\
      &=& 
\frac{1}{(1 - u_1^3) (1 + 2u_1^3)}
+ 
\frac{1}{(1 + u_2^3) (2u_2^3 - 1)} . 
      \label{eqn:a}
 \end{eqnarray}

\begin{figure}
\includegraphics[angle=0,scale=0.8]{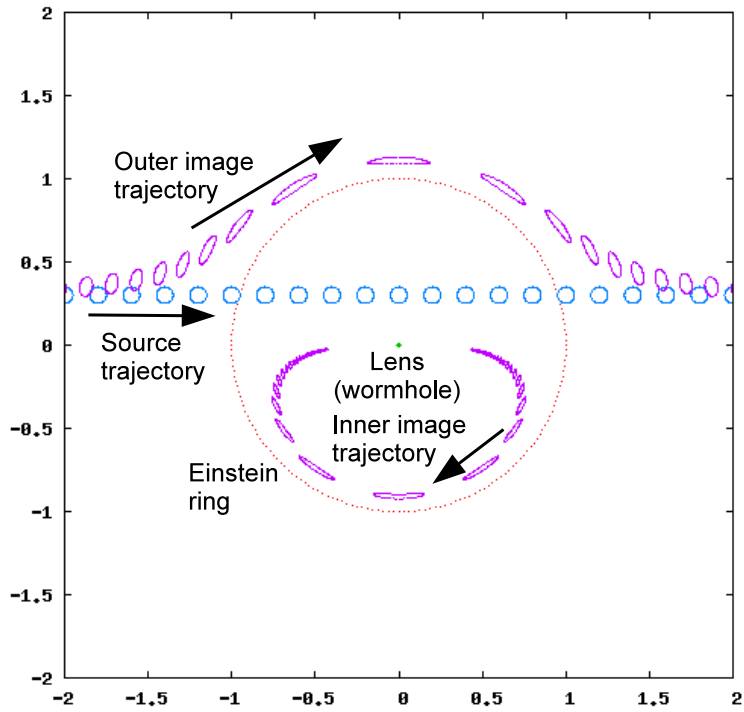}
\caption{Source and image trajectories in the sky from the position 
of the observer.\label{fig2}}
\end{figure}

\begin{figure}
\includegraphics[angle=0,scale=0.6]{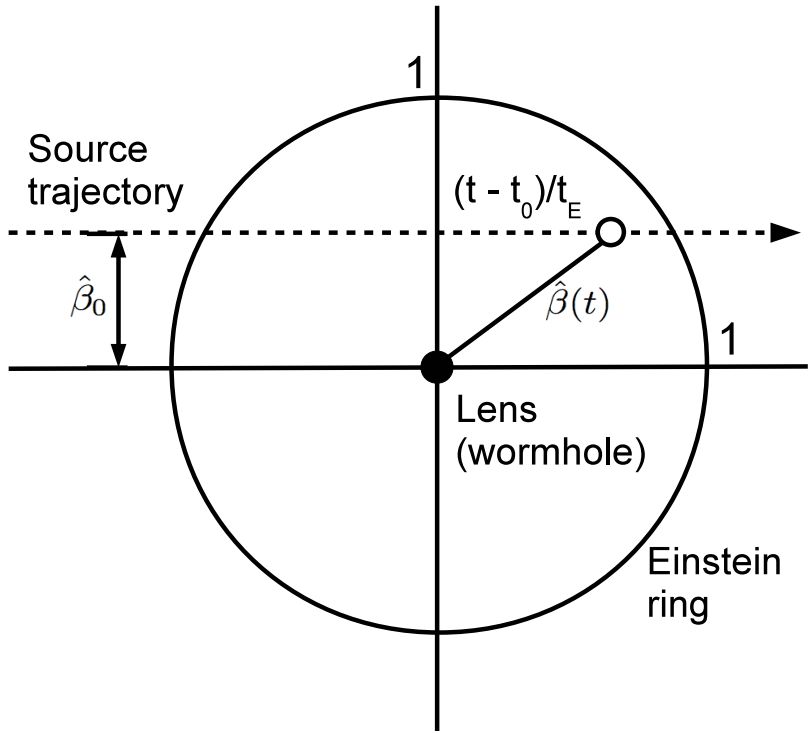}
\caption{Sketch of the relation between the source trajectory and the lens (wormhole) in the sky. All quantities are normalized by the angular Einstein radius $\theta_E$.\label{fig3}}
\end{figure}

\begin{figure}
\includegraphics[angle=0,scale=0.45]{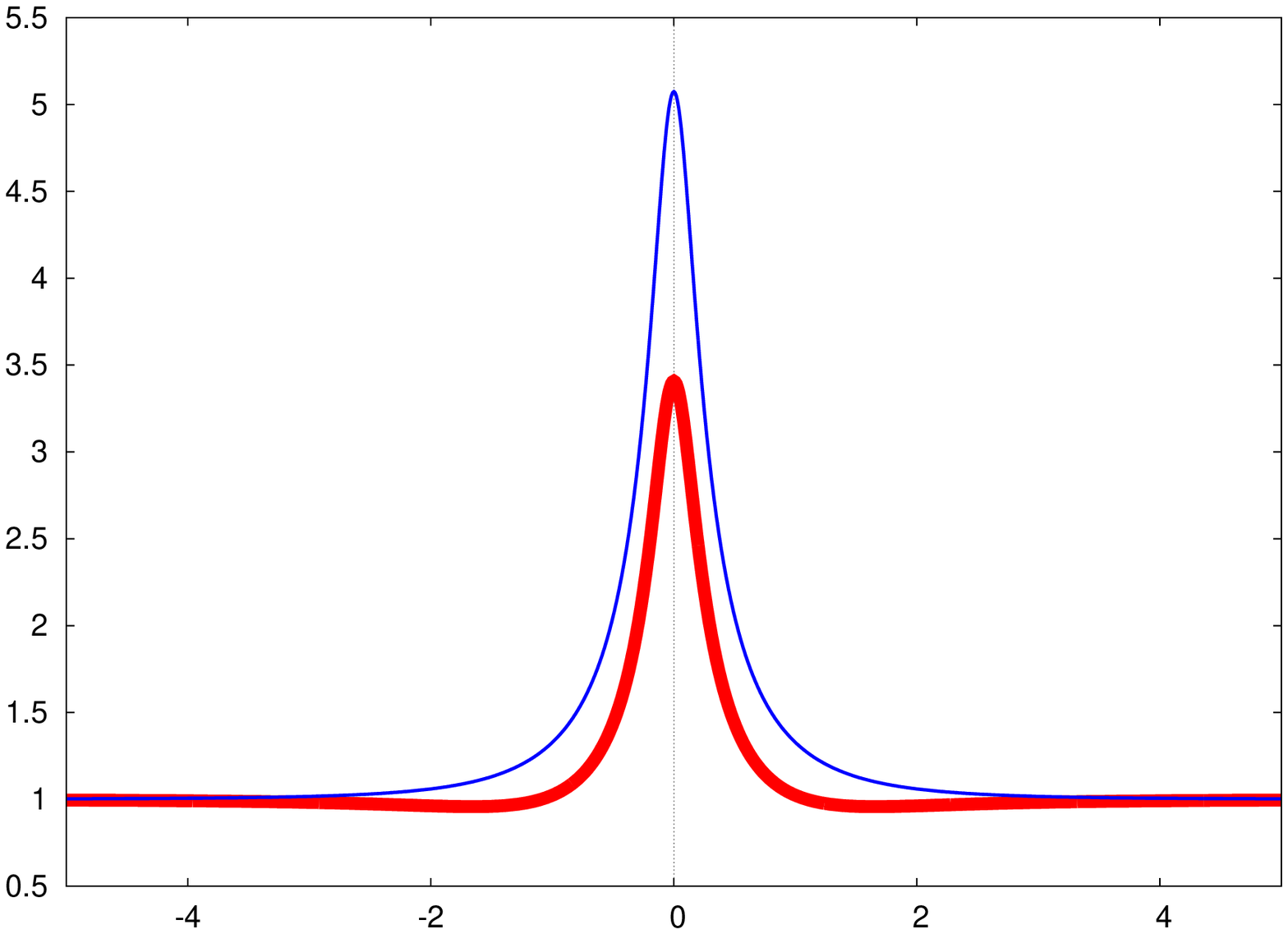}
\includegraphics[angle=0,scale=0.45]{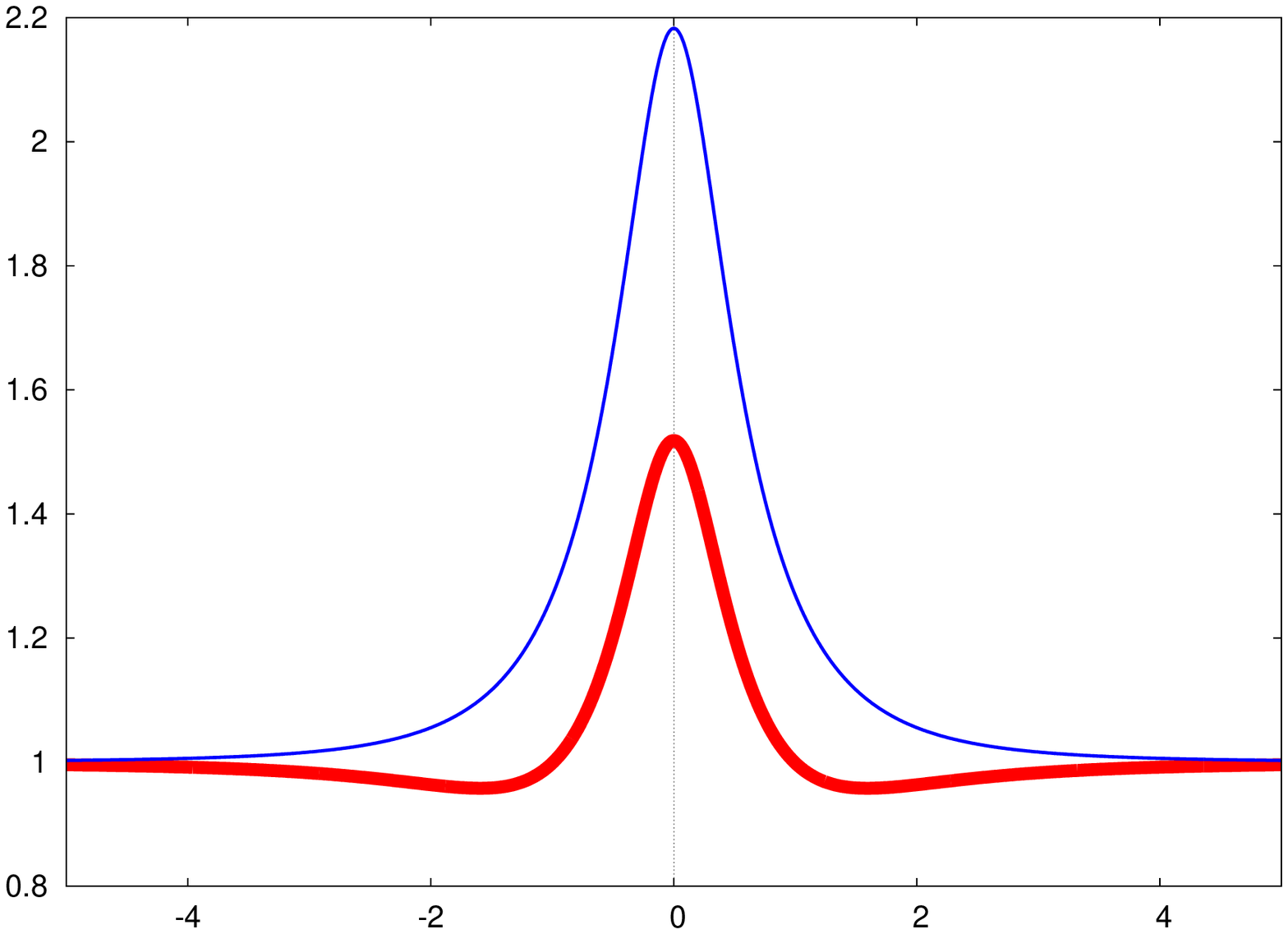}
\includegraphics[angle=0,scale=0.45]{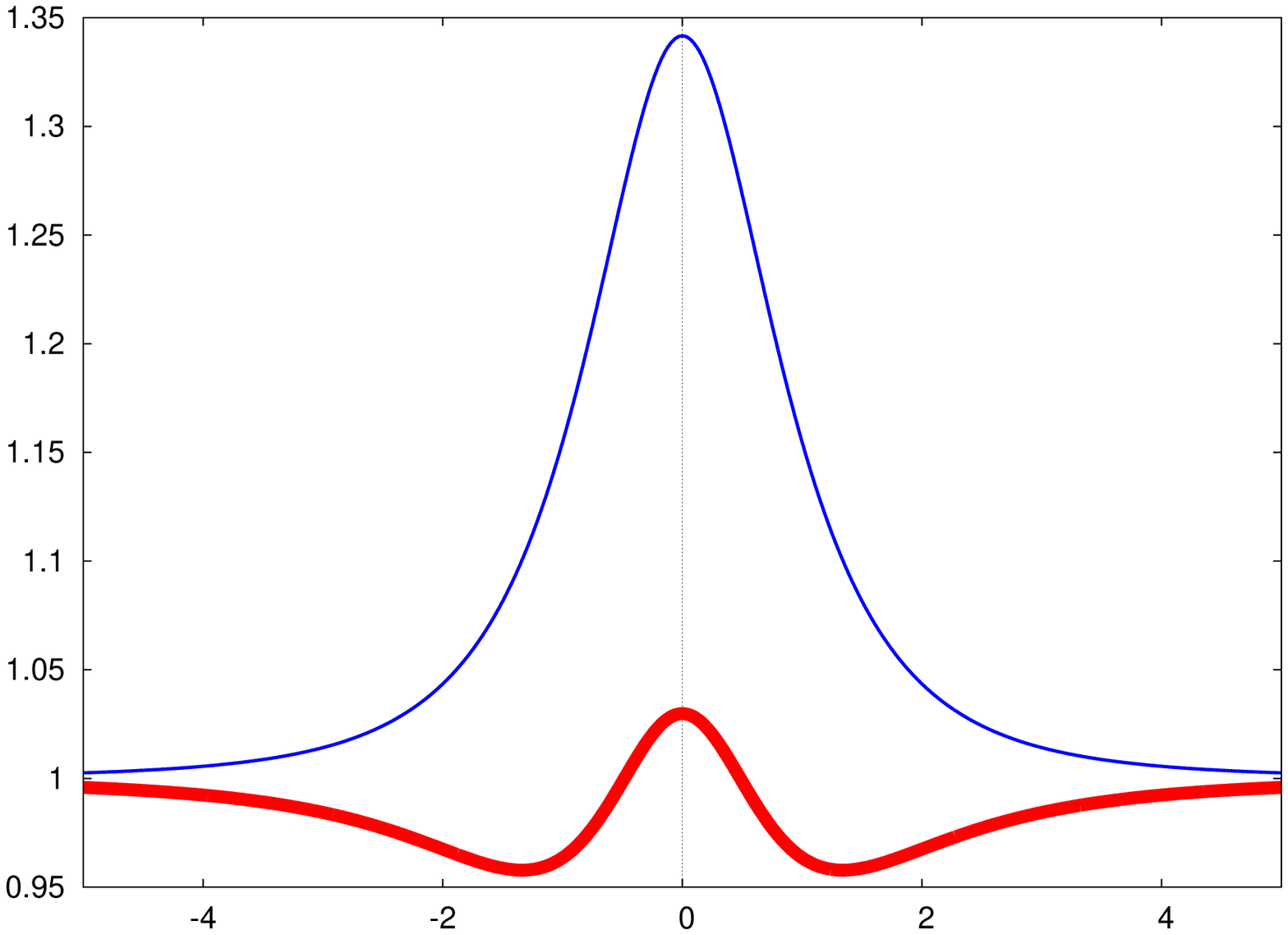}
\includegraphics[angle=0,scale=0.45]{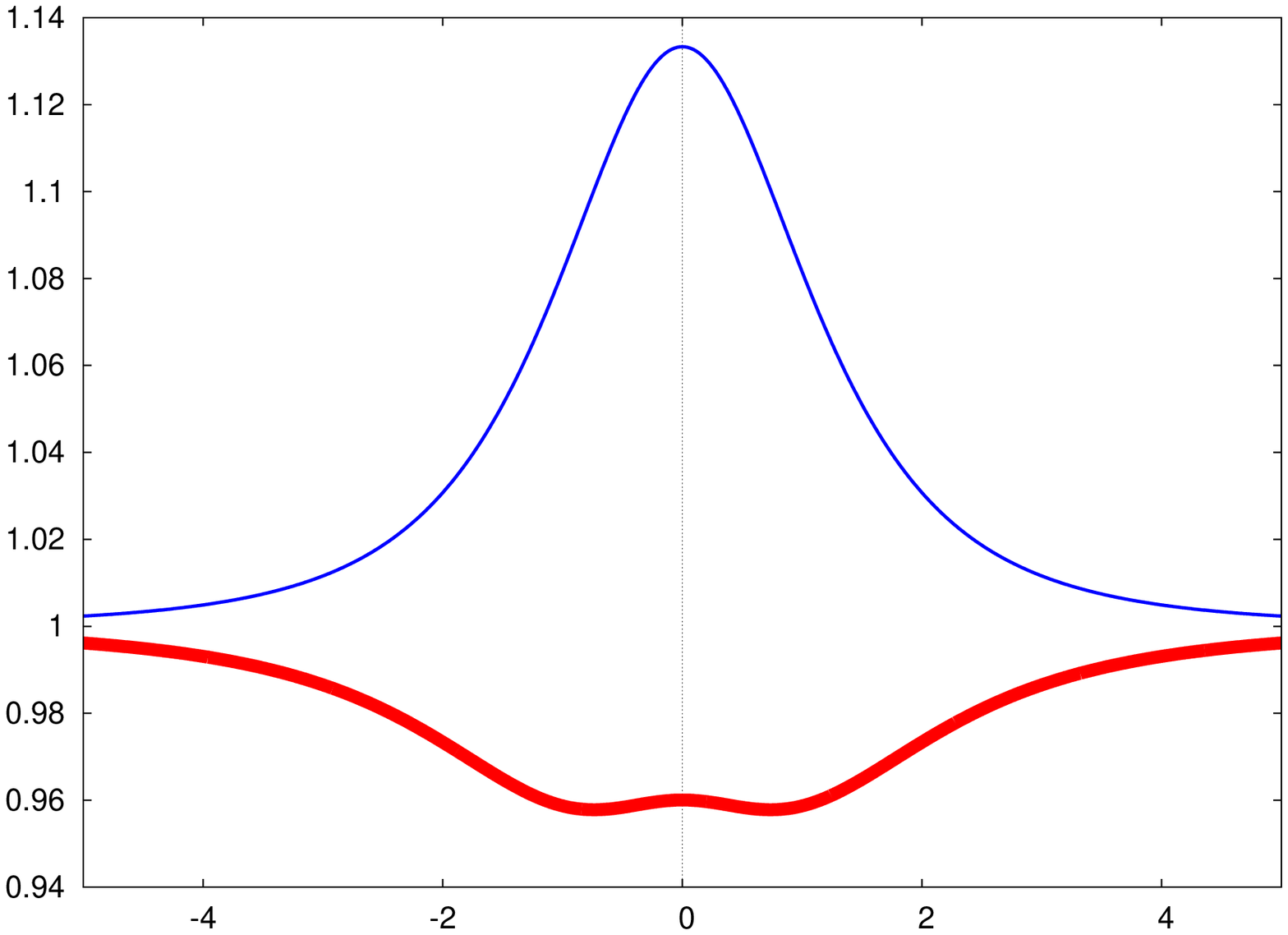}
\caption{
Light curves for 
$\hat{\beta}_0 = 0.2$ (top left), 
$\hat{\beta}_0 = 0.5 $ (top right), 
$\hat{\beta}_0 = 1.0$ (bottom left), 
and $\hat{\beta}_0 = 1.5$ (bottom right). 
Thick red lines are for wormholes. 
Thin blue lines are corresponding light curves 
for Schwarzschild lenses.
For light curves, 
the horizontal axis denotes time in units of 
the Einstein radius crossing time and 
the vertical one denotes the total magnification. 
\label{fig4}}
\end{figure}

The relation between the lens and source trajectory in the sky is 
shown in Figures \ref{fig2} and \ref{fig3}. 
The time dependence of $\hat{\beta}$ is
\begin{equation}
\hat{\beta}(t) = \sqrt{\hat{\beta}_0^2 + {(t -t_0)^2/t_E}^2}, \label{eqn:hbeta}
\end{equation}
where $\hat{\beta}_0$ is the impact parameter of the source trajectory and $t_0$ is the time of closest approach. $t_E$ is the Einstein radius crossing time given by
\begin{equation}
t_E = R_E / v_T,  \label{eqn:te}
\end{equation}
where $v_T$ is the transverse velocity of the lens relative to the
source and observer. The light curves obtained from Equations
(\ref{eqn:a}) and (\ref{eqn:hbeta}) are shown as thick red lines in
Figure \ref{fig4}. The light curves corresponding to Schwarzschild
lensing are shown as thin blue lines for comparison. 
\cite{abe10} found that 
the magnifications by the Ellis wormhole are generally less than those of
Schwarzschild lensing. 
The light curve of the Ellis wormhole for $\hat{\beta_o} < 1.0$ shows 
characteristic gutters on both sides of the peak immediately 
outside the Einstein ring crossing times ($t = t_0 \pm t_E$). 
The depth of the gutters is about 4\% from the baseline. 
Amazingly, the star becomes fainter than normal in terms of 
apparent brightness in the gutters. This means that 
the Ellis wormhole lensing has off-center divergence. 
In conventional gravitational lensing theory \citep{sch92}, 
the convergence of light is expressed by a convolution of 
the surface mass density. Thus, we need to introduce negative mass 
to describe 
a diverging lens (like a concave lens in optics) 
by the Ellis wormhole. 
However, negative mass is not a physical entity. 
Since the lensing by the Ellis wormhole is 
converging at the center, 
lensing at some other place must be diverging
because the wormhole has zero asymptotic mass 
and hence converging and diverging effects are compensated each other
in total. 

For $\hat{\beta_o} > 1.0$, the light curve 
of the wormhole has a basin at $t_0$ and no peak. 
Using these features, discrimination from Schwarzschild lensing 
can be achieved. Equations (\ref{eqn:re}) and (\ref{eqn:te}) indicate 
that physical parameters ($D_L$, $a$, and $v_T$) are degenerate 
in $t_E$ and cannot be derived by fitting the light-curve data. 
This situation is the same as that for Schwarzschild lensing. 
To obtain or constrain these values, observations of 
the finite-source effect \citep{nem94} or parallax \citep{alc95} 
are necessary.
Astrometry gives a method for breaking the degeneracy 
as discussed later.

\begin{figure}
\includegraphics[angle=0,scale=0.45]{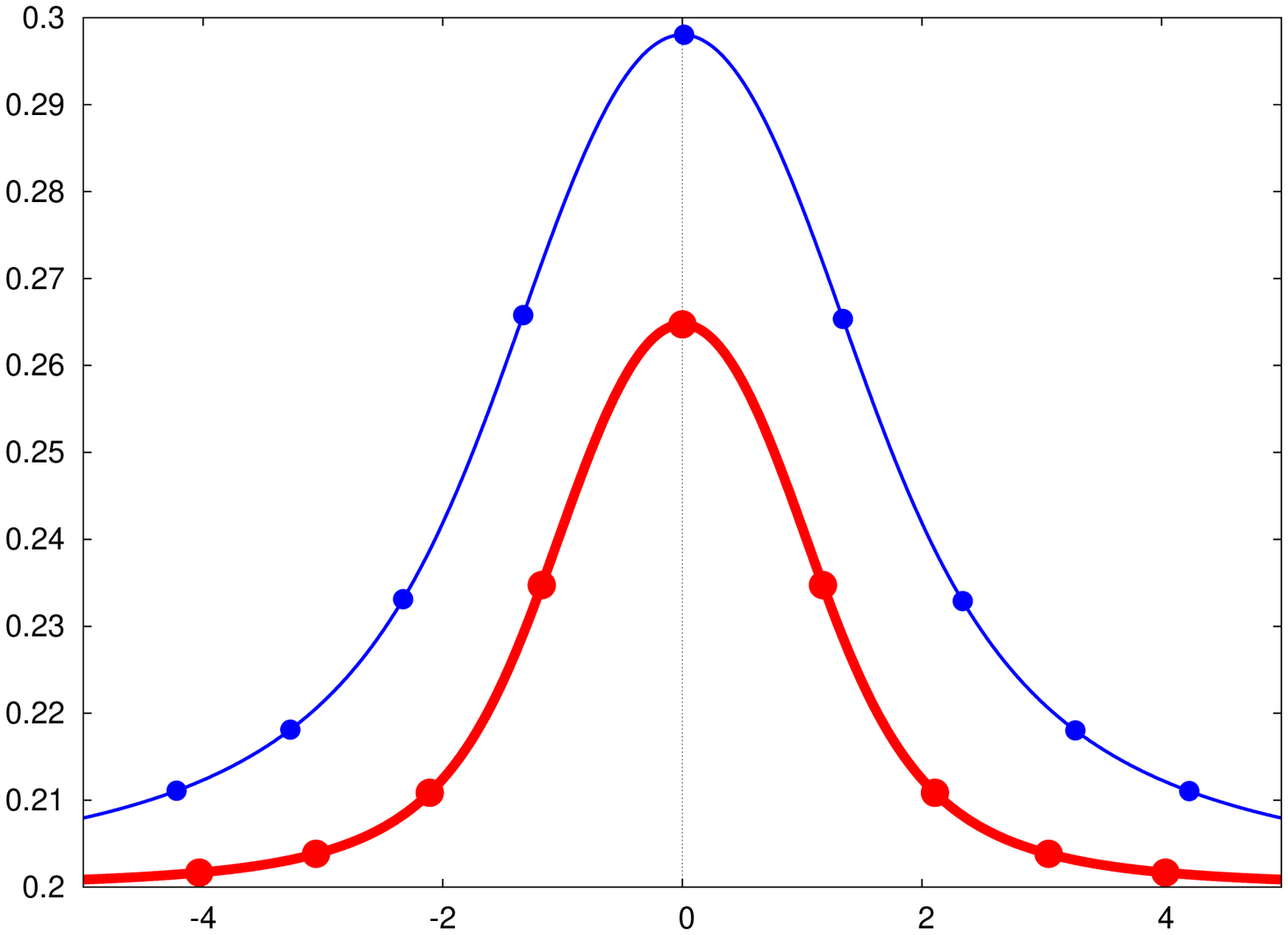}
\includegraphics[angle=0,scale=0.45]{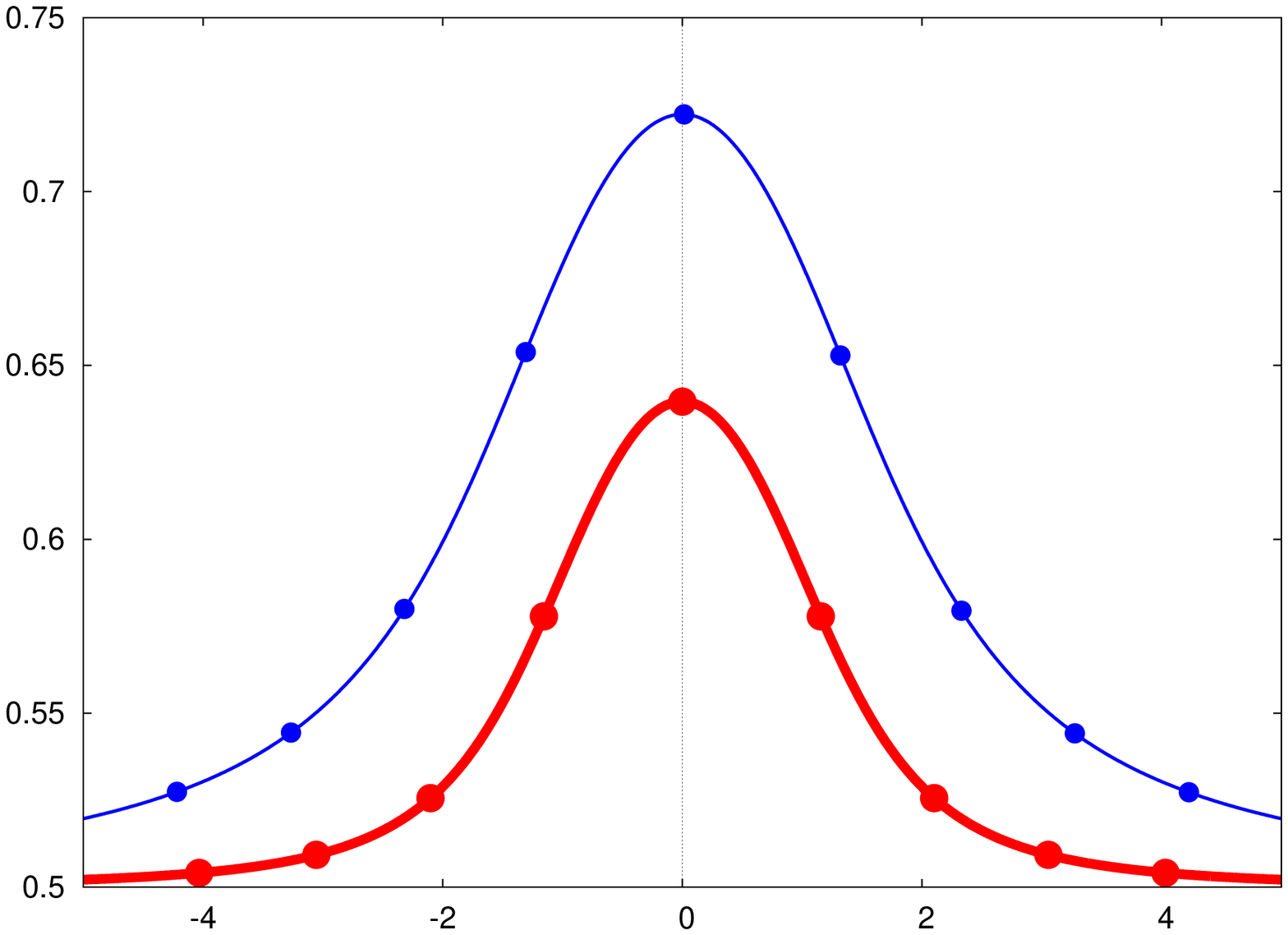}
\includegraphics[angle=0,scale=0.45]{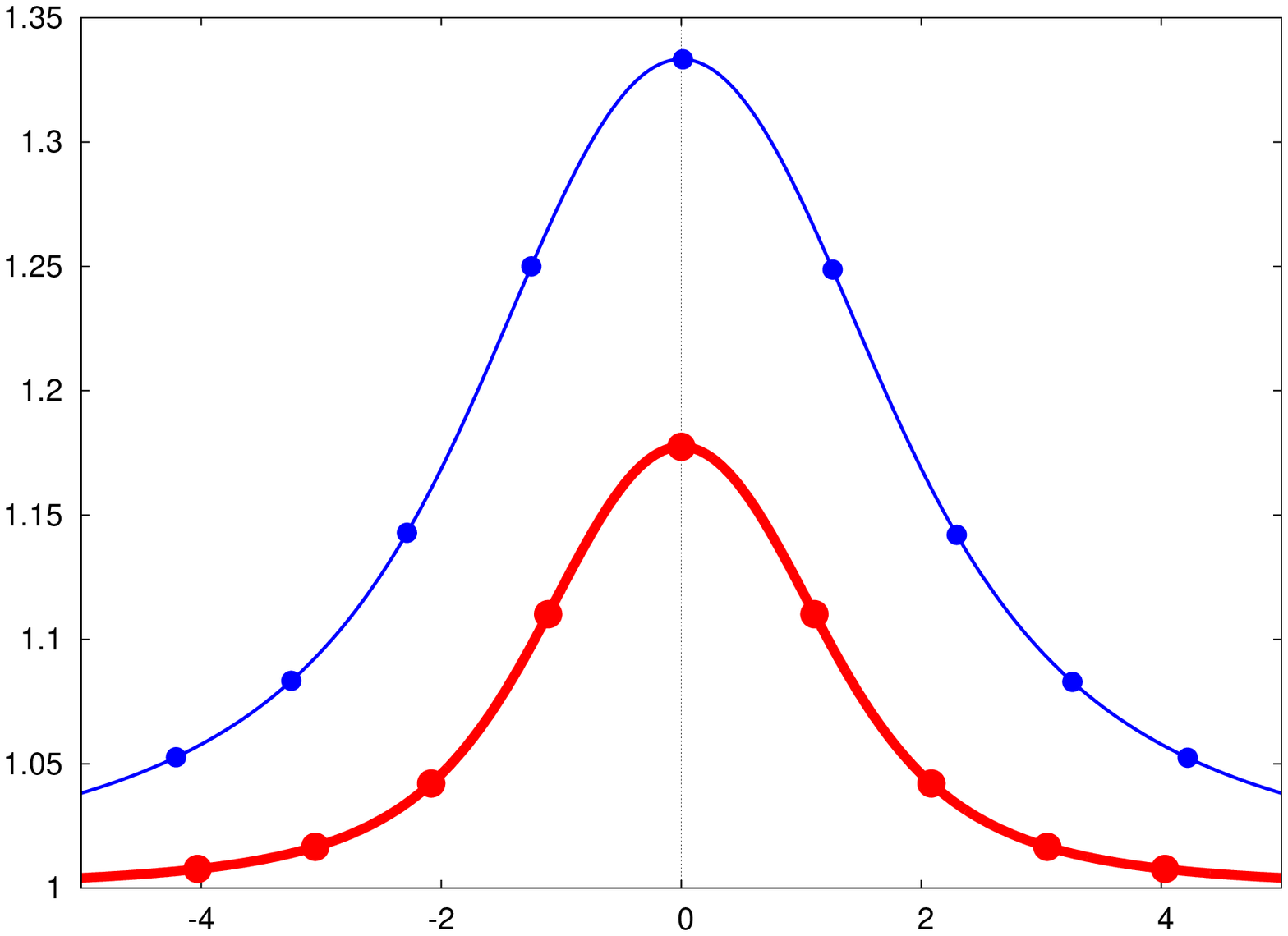}
\includegraphics[angle=0,scale=0.45]{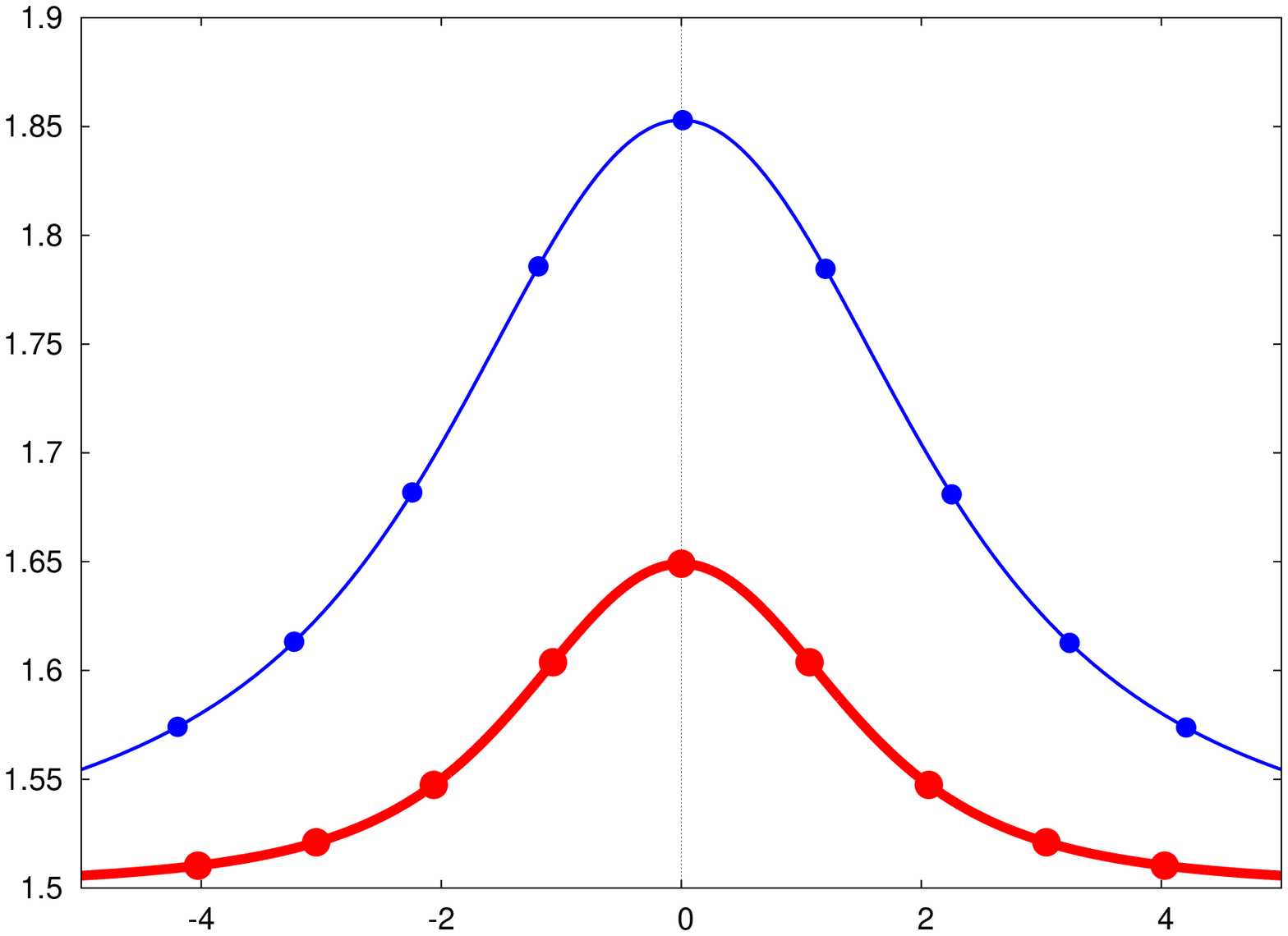}
\caption{ 
Image centroid trajectories 
both by the Ellis wormhole lens 
and by the Schwarzschild one 
with the same Einstein ring radius 
for 
$\hat{\beta}_0 = 0.2$ (top left), 
$\hat{\beta}_0 = 0.5 $ (top right), 
$\hat{\beta}_0 = 1.0$ (bottom left), 
and $\hat{\beta}_0 = 1.5$ (bottom right). 
Thick red lines are image centroid orbits 
expressed as $(\hat\theta_{pc,x}(t), \hat\theta_{pc,y}(t))$ for wormholes. 
Thin blue lines correspond to 
Schwarzschild lens cases.
The horizontal axis $(\hat\theta_{pc,x})$ is taken along the source motion 
and the vertical one $(\hat\theta_{pc,y})$ is normal to the direction
of its motion. 
Scales are normalized by the Einstein ring radius, 
where we use Eq. (\ref{pc}). 
In order to make the image centroid motion clear, 
we plot the image centroid position at $t=-4, -3, -2, -1, 0, 1, 2, 3, 4$, 
which are marked as filled disks in the figure. 
For the Ellis wormhole case, 
the lensed centroid moves slower
than that for the Schwarzschild one, 
provided that the Einstein ring radius and 
the impact parameter are the same.  
\label{fig5}}

\end{figure}

In analogy with the center of the mass distribution, 
the centroid position of the light distribution of 
a gravitationally microlensed source is given by 
\begin{eqnarray}
\hat{\theta}_{pc} 
&=& 
\frac{A_{1} \hat{\theta}_{1} + A_{2} \hat{\theta}_{2}}{A} . 
\label{pc}
\end{eqnarray}
In making numerical figures, we employ $x-y$ coordinates 
in the way that the center is chosen as the location of the Ellis wormhole, 
$x$-axis is taken along the direction of the source motion 
and $y$-axis is perpendicular to the source motion.  
See Figure \ref{fig5} for 
the image centroid trajectories for 
$\hat\beta_0 = 0.2, 0.5, 1.0, 1.5$. 
For each $\hat\beta_0$, 
the maximum difference between the image centroid position 
by the Ellis wormhole and that by the Schwarzschild one is  
$-0.03, -0.08, -0.16, -0.20$ 
in the units of the Einstein ring radius, 
respectively. 
This implies that the astrometric lensing by the Ellis wormhole 
is relatively weaker than that by the Schwarzschild one. 

\begin{figure}
\includegraphics[angle=0,scale=0.45]{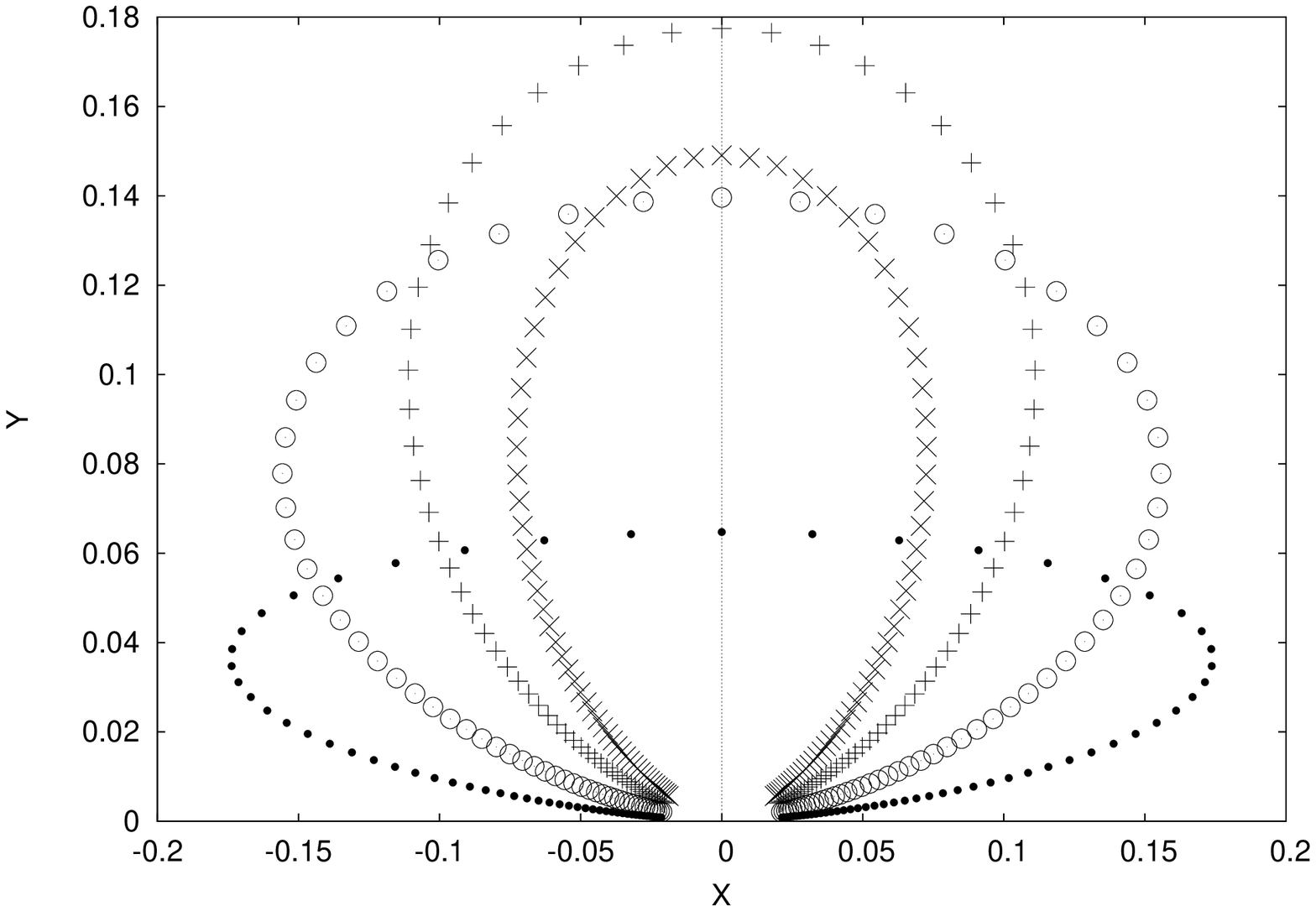}
\includegraphics[angle=0,scale=0.45]{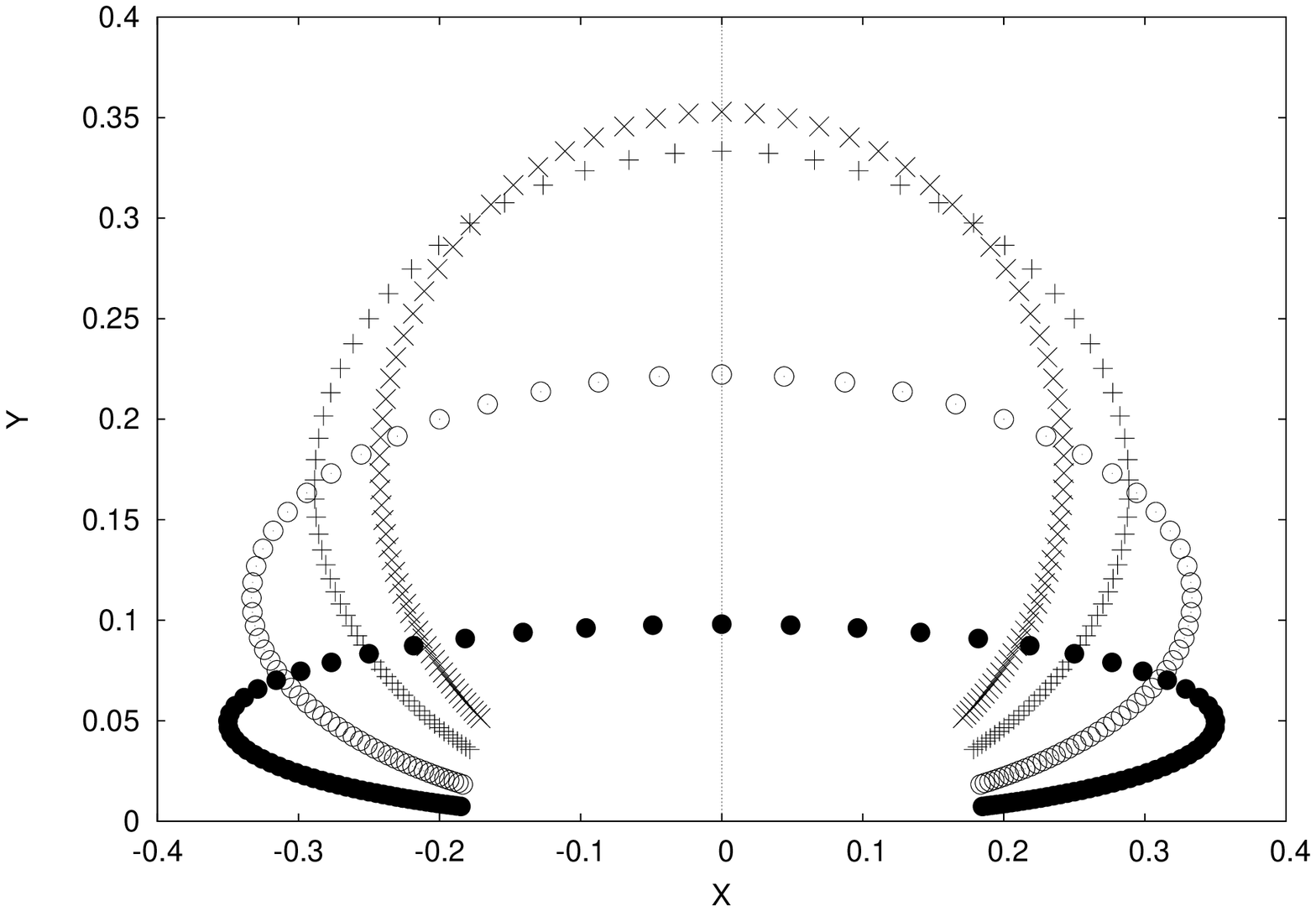}
\caption{
Relative displacements of the image centroid   
by the Ellis wormhole lens (left panel) 
and by the Schwarzschild one (right panel) 
with respect to the source position in uniform linear motion. 
The parameter values (normalized by the Einstein ring radius) 
are the same as those in Figure \ref{fig5}. 
The dot ($\cdot$), circle ($\bigcirc$), plus ($+$), cross ($\times$) 
denote the positions 
for $\hat\beta_0 = 0.2, 0.5, 1.0, 1.5$, respectively, 
where the horizontal axis denotes the direction of the source motion. 
Points are spaced uniformly in time as 
$\Delta t = 0.1$. 
In the limit of $t = \pm \infty$, these points approach 
the origin of the $X-Y$ plane, 
namely the image centroid position agrees with the source. 
Hence, points around the origin are very crowded and thus we  
omit the crowded points by choosing the time domain from $t=-20$ to $20$. 
\label{fig6}}
\end{figure}

In the weak-field region, 
the suppression of the anomalous shift of the image centroid position 
is because the bending angle by the Ellis wormhole 
is proportional to the inverse squared impact parameter, 
whereas that by the Schwarzschild lens depends on 
the inverse impact parameter. 
Figure \ref{fig6} shows the relative displacement  
of the image centroid with respect to the source position 
that is assumed to be in uniform linear motion. 
The maximum vertical displacement is 
$0.06, 0.14, 0.18, 0.15$ 
for $\hat\beta_0 = 0.2, 0.5, 1.0, 1.5$, 
respectively. 
Here, a key question is whether 
the Ellis lensing and the Schwarzschild one are distinguished from 
the centroid displacement curve. 
The relative displacement trajectory by the Schwarzschild lens 
is known to be an ellipse \citep{Walker,JHP}. 
It is natural to ask whether the displacement curve 
by the Ellis wormhole lens is also an ellipse. 
Figure \ref{fig6} shows that 
the relative trajectory by the Ellis lens 
looks like an ellipse but has a small difference. 
The shape is symmetric along the $x$-axis but 
slightly asymmetric along the $y$-axis 
like a tree leaf, 
particularly for $\hat\beta_0 = 0.2$. 
Figure \ref{fig6}, however, shows that 
such a deviation of the relative trajectory from elliptic orbits 
is very small. 
Another difference is that the relative displacement 
at large $t$, for instance $t=-20$ or $20$, 
is dependent strongly on the Ellis lens or the Schwarzschild one. 
This is because the asymptotic behavior of the centroid 
displacement is different 
($\hat\beta^{-2}$ or $\hat\beta^{-1}$). 
In other words, the displacement effect by the Ellis lens 
goes away faster.  
This suggests that a long-term observation including a tail part 
of the centroid curve is required to distinguish 
the Ellis lenses by astrometric observations alone.

The detectability of the image centroid displacements of the background star 
depends on the timescale called 
{\it the Einstein radius crossing time} $t_E$ 
that depends on the transverse velocity $v_T$. 
There is no reliable estimate of $v_T$ for wormholes. 
Following Abe (2010), 
we assume that the velocity of the wormhole is approximately 
equal to the rotation velocity of stars ($v_T = 220 km/s$) 
if it is bound to the Galaxy. 
If the wormhole is not bound to our Galaxy, 
the transverse velocity would be much higher. 
We assume $v_T = 5000 km/s$ \citep{saf01} for the unbound wormhole. 
Table 2 shows the Einstein radius crossing times of 
the Ellis wormhole lensings for the Galactic bulge and LMC 
in both bound and unbound scenarios. 
As the frequencies of current microlensing observations are limited 
to once every few hours, an event for which the timescale is 
less than one day is difficult to detect. 
To find very long timescale events ($t_E \geq 1000 days$), 
long-term monitoring of events is necessary. 
The realistic period of observation is $\leq 10 years$. 
Thus, the realistic size of the throat that we can search 
for is limited to $100 km \le a \le 10^7 km$ both 
for the Galactic bulge and LMC if wormholes are bound to our Galaxy. 
If wormholes are unbound, the detection is limited to 
$ 10^5 km \le a \le 10^9 km$. 

The detectability depends also on the angular shift 
due to the Ellis wormhole lens. 
The typical angular scale is $O(\theta_E)$. 
See Table 1 for the size of $\theta_E$ 
corresponding to various values of the throat radius. 
Near future astrometry space missions such as Gaia 
and JASMINE are expected to have angular sensitivity 
of a few micro arcseconds, 
for which the detection is limited as 
$a \geq 10^2$ km. 
This limit is much weaker than that by 
the mission life time as 
$10^5 \mbox{km} \leq a \leq 10^9 \mbox{km}$ 
for unbound wormholes. 

Note that there is a small difference in the image centroid position 
(and its motion with time) between the Scwarzschild lensing and 
the Ellis wormhole one. 
In practice, therefore, it is unlikely to detect 
the wormhole by astrometric observation alone. 
It is safe to say that 
the astrometric lensing provides a supplementary method 
of supporting a photometric detection: 
First, the impact parameter of the source trajectory $\hat\beta_0$ 
is determined from light curve observations. 
By using the obtained $\hat\beta_0$, 
one can fit the astrometric observations 
with wormhole lensing templates. 
If astrometric data show a better fit with a wormhole case, 
the detection by light curves will be reinforced. 
What is more important is that 
astrometric observations give an additional information 
such as the angular size of the image centroid position shift, 
so that the degeneracy among $(D_L, a, v_T)$ can be partially broken. 

Before closing this section, let us briefly summarize 
the chain of logic for identifying Ellis lenses. 
{}From light curves, first, 
we distinguish the Ellis lenses from the Schwarzschild ones. 
The best fit values of the model parameter combinations 
are obtained from them. 
Next, the image centroid observations are used 
to partially break the parameter degeneracy.

\section{Summary}
We studied the gravitational microlensing 
effects of the Ellis wormhole in the weak-field limit. 
First, we performed a suitable coordinate transformation, 
such that the lens equation 
and analytic expressions of the lensed image positions 
can become much simpler than the previous ones. 
Second, we proved that two images always appear 
for the weak-field lens by the Ellis wormhole. 
By using these analytic results, 
we investigated astrometric image centroid displacements due to 
gravitational microlensing by the Ellis wormhole. 
The anomalous shift of the image centroid by the Ellis wormhole lens 
is smaller than that by the Schwarzschild lens, 
provided that the impact parameter and the Einstein ring radius 
are the same. 
Therefore, the lensed image centroid by the Ellis wormhole 
moves slower. 

Studies of astrometric image centroid displacements 
due to another type of wormholes or nontrivial topology 
of spacetimes are left as a future work.

\acknowledgments
We would like to thank Professor Volker Perlick for invaluable 
comments on the Ellis wormhole lensing in 
the strong-field region. 
This work was supported in part (H.A.) 
by a Japanese Grant-in-Aid 
for Scientific Research from the Ministry of Education, 
No. 21540252.









\begin{center}
\begin{table}
\caption{Einstein radii for bulge and LMC lensings\label{tbl-1}} 
\begin{tabular}{rrrrrrr}
\hline \hline
  & & \multicolumn{2}{c}{Bulge\tablenotemark{a}} & &  \multicolumn{2}{c}{LMC\tablenotemark{b}} \\ \cline{3-4} \cline{6-7}
  $a (km)$ & &  $R_E (km)$ & $\theta_E (mas)$ & & $R_E (km)$ & $\theta_E (mas)$ \\
 \hline
1            & & $3.64 \times 10^5$ & 0.001 & & $6.71 \times 10^5$  & $< 0.001$ \\
10          &  & $1.69 \times 10^6$ & 0.003 & & $3.12 \times 10^6$ & 0.001 \\
$10^2$ &  & $7.85 \times 10^6$ & 0.013 & & $1.45 \times 10^7$ & 0.004 \\
$10^3$ &  & $3.64 \times 10^7$ & 0.061 & & $6.71 \times 10^7$ & 0.018 \\
$10^4$ &  & $1.69 \times 10^8$ & 0.283 & & $3.12 \times 10^8$ & 0.083  \\
$10^5$ &  & $7.85 \times 10^8$ & 1.31    & & $1.45 \times 10^9$ & 0.387  \\
$10^6$ &  & $3.64 \times 10^9$ & 6.10    & &  $6.71 \times 10^9$ &  1.80 \\
$10^7$ &  & $1.69 \times 10^{10}$ & 28.3    & &  $3.12 \times 10^{10}$ &  8.35 \\
$10^8$ &  & $7.85 \times 10^{10}$ & 131    & &  $1.45 \times 10^{11}$ &  38.7 \\
$10^9$ &  & $3.64 \times 10^{11}$ & 610    & &  $6.71 \times 10^{11}$  &  180 \\
$10^{10}$ & & $1.69 \times 10^{12}$ & 2 832    & &  $3.12 \times 10^{12}$ &  835 \\
$10^{11}$ & & $7.85 \times 10^{12}$ & 13 143    & &  $1.45 \times 10^{13}$ &  3 874 \\
\hline \hline
\end{tabular}
\tablecomments{$a$ is the throat radius of the wormhole, $R_E$ is the Einstein radius, and $\theta_E$ is the angular Einstein radius.}
\tablenotetext{a}{$D_S = 8 kpc$ and $D_L = 4 kpc$ are assumed. }
\tablenotetext{b}{$D_S = 50 kpc$ and $D_L = 25 kpc$ are assumed. }
\end{table}
\end{center}


\begin{center}
\begin{table}
\caption{Einstein radius crossing times for bulge and LMC lensings\label{tbl-2}}
\begin{tabular}{rrrrrrrrr}
\hline \hline
  & & \multicolumn{2}{c}{Bulge\tablenotemark{a}} & & \multicolumn{2}{c}{LMC\tablenotemark{b}} \\ 
  $a (km)$ & & \multicolumn{2}{c}{$t_E (day)$} & & \multicolumn{2}{c}{$t_E (day)$}  \\ \cline{3-4} \cline{6-7}
 & & Bound\tablenotemark{c} & Unbound\tablenotemark{d} & & Bound\tablenotemark{c} & Unbound\tablenotemark{d} \\
 \hline
1            & & 0.019 & 0.001 & & 0.035 & 0.002   \\
10          & & 0.089 & 0.004 & & 0.164 & 0.007  \\
$10^2$ & & 0.413 & 0.018 & & 0.761 & 0.033  \\
$10^3$ & & 1.92   & 0.084 & & 3.53   & 0.155  \\
$10^4$ & & 8.90    & 0.392 & &16.4   & 0.721  \\
$10^5$ & & 41.3    & 1.82   & & 76.1   & 3.35  \\
$10^6$ & & 192    & 8.44    & & 353   & 15.5   \\
$10^7$ & & 890    & 39.2  & & 1 639   & 72.1 \\
$10^8$ & & 4 130    & 182    & & 7 608   & 335  \\
$10^9$ & & $> 10^4$    & 843    & & $> 10^4 $  & 1 553  \\
$10^{10}$ & & $> 10^4 $  & 3915    & & $> 10^4$  & 7 212 \\
\hline \hline
\end{tabular}
\tablecomments{$a$ is the throat radius of the wormhole, $t_E$ is the Einstein radius crossing time.}
\tablenotetext{a}{$D_S = 8 kpc$ and $D_L = 4 kpc$ are assumed. }
\tablenotetext{b}{$D_S = 50 kpc$ and $D_L = 25 kpc$ are assumed. }
\tablenotetext{c}{$v_T = 220 km/s$ is assumed.}
\tablenotetext{d}{$v_T = 5000 km/s$ is assumed.}
\end{table}
\end{center}







\end{document}